\documentclass[11pt]{article}
\usepackage{exscale}\usepackage{graphicx}
\usepackage{amssymb,amsmath}

\def\d{\mbox{\rm d}}
\def\e{\mbox{\rm e}}

\def\dddot#1{\mathinner{\buildrel\vbox{\kern5pt\hbox{...}}\over{#1}}}
\def\ddddot#1{\mathinner{\buildrel\vbox{\kern5pt\hbox{....}}\over{#1}}}

\begin{document}
\title{An old Method of Jacobi to find Lagrangians}
 \author{M.C. Nucci  $\;$ and $\;$ P.G.L. Leach\footnote{permanent address: School of Mathematical Sciences, Westville Campus,
University of KwaZulu-Natal, Durban 4000, Republic of South
Africa, e-mail: leachp@ukzn.ac.za, leachp@math.aegean.gr}}
\date{Dipartimento di Matematica
e Informatica,Universit\`a di Perugia, 06123 Perugia, Italy,
e-mail: nucci@unipg.it}

 \maketitle

\begin{abstract}
In a recent paper by Ibragimov [N. H. Ibragimov, Invariant
Lagrangians and a new method of integration of nonlinear
equations, J. Math. Anal. Appl. 304 (2005) 212--235] a method was
presented in order to find Lagrangians of certain second-order
ordinary differential equations admitting a two-dimensional Lie
symmetry algebra. We present a method devised by Jacobi which
enables to derive (many) Lagrangians of any second-order
differential equation. The method is based on the search of the
Jacobi Last Multipliers of the equations. We exemplify the
simplicity and elegance of Jacobi's method by applying it to the
same two equations as did Ibragimov. We show that the Lagrangians
obtained by Ibragimov are particular cases of some of the many
Lagrangians that can be obtained by Jacobi's method.
\end{abstract}

\section {Introduction}

The method of the Jacobi last multiplier
 \cite{Jacobi 42 a,Jacobi 44 a,Jacobi 44 b,Jacobi 45 a,Jacobi 86 a}
  provides a means to determine an
integrating factor of the partial differential equation
\begin {equation}
Af = \sum_{i = 1} ^n a_i(x_1,\dots,x_n)\frac {\partial f} {\partial x_i} = 0 \label {12.1}
\end {equation}
or its equivalent associated Lagrange's system
\begin {equation}
\frac {\d x_1} {a_1} = \frac {\d x_2} {a_2} = \ldots = \frac {\d x_n} {a_n}.\label {12.2}
\end {equation}
The multiplier $M$ is given by
\begin {equation}
\frac {\partial (f,\omega_1,\omega_2,\ldots,\omega_{n- 1})}{\partial (x_1,x_2,\ldots,x_n)}  = MAf, \label {12.3}
\end {equation}
where
\begin {equation}
\frac {\partial (f,\omega_1,\omega_2,\ldots,\omega_{n- 1})}{\partial (x_1,x_2,\ldots,x_n)} = \mbox {\rm det}\left [
\begin {array} {ccc}
\displaystyle {\frac {\partial f} {\partial x_1}} &\cdots &\displaystyle {\frac {\partial f} {\partial x_n}}\\
\displaystyle {\frac {\partial\omega_1} {\partial x_1}} & &\displaystyle {\frac {\partial\omega_1} {\partial x_n}}\\
\vdots & &\vdots\\
\displaystyle {\frac {\partial\omega_{n- 1}} {\partial x_1}} &\cdots &\displaystyle {\frac {\partial\omega_{n- 1}} {\partial x_n}}
\end {array}\right] = 0 \label {12.4}
\end {equation}
and $\omega_1,\ldots,\omega_{n- 1} $ are $n- 1 $ solutions of
(\ref {12.1}) or, equivalently, first integrals of (\ref {12.2}).
Jacobi also proved that $M$ is a solution of the following linear
partial differential equation
\begin {equation}
\sum_{i = 1} ^n \frac {\partial (Ma_i)} {\partial x_i} = 0. \label{12.5}
\end {equation}
In general a different selection of integrals produces another
multiplier, $\tilde M$. An important property of the last
multiplier is that the ratio, $M/\tilde M$, is a solution of (\ref
{12.1}), equally a first integral of (\ref {12.2}). Indeed, if
each component of the vector field of the equation of motion is
free of the variable associated with that component, {\it ie}
$\partial a_i/\partial
x_i = 0 $, the last multiplier is a constant.\\
In its original formulation the method of Jacobi last multiplier
required almost complete knowledge of the system, (\ref {12.1}) or
(\ref {12.2}), under consideration\footnote{Although we should
underline that Jacobi himself found last multipliers for several
equations without any knowledge of its solutions \cite{Jacobi 42
a,Jacobi 44 a,Jacobi 44 b,Jacobi 45 a,Jacobi 86 a}.}. Since the
existence of a solution/first integral is consequent upon the
existence of symmetry, an alternative formulation in terms of
symmetries was provided by Lie \cite {Lie 74 a,Lie 12 a}[Kap 15,
\S 5 in the latter].  A clear treatment of the formulation in
terms of solutions/first integrals and symmetries is given by
Bianchi \cite {Bianchi 18 a}. If we know $n- 1 $ symmetries of
(\ref {2.1})/(\ref {2.2}), say
\begin {equation}
\Gamma_i = \sum_{j=1}^{n}\xi_{ij}(x_1,\dots,x_n)\partial_{x_j},\quad i = 1,n-1, \label {12.6}
\end {equation}
Jacobi Last Multiplier is given by $M =\Delta ^ {- 1} $, provided
that $\Delta\not = 0 $, where
\begin {equation}
\Delta = \mbox {\rm det}\left [\begin {array} {ccc}
a_1 &\cdots & a_n\\
\xi_{1,1} & &\xi_{1,n}\\
\vdots & &\vdots\\
\xi_{n- 1,1}&\cdots &\xi_{n- 1,n}
\end {array}\right]. \label {12.8}
\end {equation}
There is an obvious corollary to the results of Jacobi mentioned
above. In the case that there exists a constant multiplier, the
determinant $\Delta$ is a first integral.  This result is
potentially very useful in the search for first integrals of
systems of ordinary differential equations. In particular this
feature was put to good use with the Euler-Poinsot system
 \cite{Nucci 02 a} and the Kepler problem \cite{Nucci 04 a}. \\
The following relationship between the Jacobi Last Multiplier and
the Lagrangian
 \cite {Jacobi 86 a}, \cite{Whittaker 44 a}
\begin {equation}
\frac {\partial ^ 2L} {\partial{y'} ^ 2} = M \label {1.1}
\end {equation}
for a one-degree-of-freedom system
\begin {equation}y''=f(x,y,y'),\end{equation}
where the prime denotes differentiation with respect to the
independent variable $x $, is perhaps not widely known although it
is certainly not unknown as can be seen from the bibliography in
\cite {Nucci 05 a}.  Given a knowledge of a multiplier, namely a
solution of the equation (\ref{12.5}), i.e.
\begin {equation} \frac{{\rm d}}{{\rm d} x}(\log M)+\frac{\partial f}{\partial
y'} =0,\label{Meq}
\end{equation}
then (\ref {1.1}) gives a simple recipe for the generation of a
Lagrangian. The only possible difficulty is the performance of the
double quadrature. Considering the dual nature of the Jacobi Last
Multiplier as providing a means to determine both Lagrangians and
integrals one is surprised that it has not attracted more
attention over the more than one and a half centuries since its
introduction.  The bibliography of \cite {Nucci 05 a} gives a fair
indication of its significant applications in the past.  In more
recent years we have presented the application of Jacobi Last
Multiplier to many different problems \cite{Nucci 02 a,Nucci 04
a,Nucci 05 a,Nucci 05 b,Nucci
06 a,Nucci 06 b,Nucci 07 a,Nucci 07 b,Nucci 07 c}.\\
In a recent  paper Ibragimov \cite {Ibragimov 05 a} proposed a
practical approach to the resolution of the classical problem of
finding the Lagrangian given a second-order ordinary differential
equation.  In his method Ibragimov introduced the idea of an
invariant Lagrangian, and derived Lagrangians of two second-order
differential equations  after lengthy calculations involving
integration of auxiliary differential equations. For the details
of the method the interested reader should consult the paper \cite
{Ibragimov 05 a}.
 \strut\hfill

In this paper we exemplify the simplicity and elegance of the
forgotten method devised by Jacobi for finding Lagrangians by
applying it to the same two equations as did
Ibragimov\footnote{Both equations are found in the textbook
\cite{Ibragimov 99 a}.  The first is example (12.27) on page 291
and the second is Exercise 12.3 on page 300.}. Specifically we
 obtain Jacobi Last Multipliers, and therefore Lagrangians,
of the equations
\begin {eqnarray}
y'' &=& \frac {y'} {y ^ 2} -  \frac {1} {xy} \label {1.7} \\
y'' &=& \e ^y - \frac {y'} {x} \label {1.8}
\end {eqnarray}
which possess the Lie point symmetries
\begin {equation}
\Gamma_1 = 2x\partial_x+ y\partial_y,\quad\quad
\Gamma_2 = x ^ 2\partial_x+ xy\partial_y \label {1.9}
\end {equation}
and
\begin {equation}
\Sigma_1 = x\partial_x - 2\partial_y,\quad\quad \Sigma_2 = x\log
(x)\partial_x - 2 \left (1+\log (x )\right)\partial_y, \label
{1.10}
\end {equation}
respectively.  We note that both symmetries in (\ref {1.9}) and in
(\ref {1.10}) generate a Lie's Type III algebra \cite{Lie 12 a},
namely a nonabelian and transitive Lie algebra \cite{Bianchi 18
a}. \strut\hfill

\section {Jacobi Last Multipliers and Lagrangians for (\ref {1.7})}

The calculation of the Jacobi Last Multiplier requires that the
differential equation under consideration be written as a system
of first-order equations.  Thus (\ref {1.7}) becomes
\begin {eqnarray}
& & u_1' =u_2 \nonumber \\
& & u_2' = \displaystyle {\frac {u_2} {u_1^2}}-\displaystyle
{\frac {1} {xu_1}}, \label {2.1}
\end {eqnarray}
with $u_1\equiv y$, and $u_2\equiv y'$. The formula (\ref {Meq})
for the last multiplier gives a nonlocal $\exp [-\int u_1 ^ {-
2}\d x] $ which is not very useful. However, we do have the route,
(\ref {12.8}), through the determinant of the vector field and the
two symmetries.  Thus we have
\begin {equation}
\Delta_{12}  =  {\rm det} \left[ {\begin{array}{ccc} 1 &
 {u_2} & {\displaystyle \frac { {u_2}}{ {u_1}
^{2}}}  - {\displaystyle \frac {1}{x {u_1}}}  \\ [2ex]
x^{2} & x {u_1} &  {u_1} - x {u_2} \\
2x &  {u_1} &  -  {u_2}
\end{array}}
 \right]  = - {\displaystyle \frac {(x {u_1} {u_2} + x -
 {u_1}^{2})(x {u_2} -  {u_1})}{ {u_1}}}
\label {2.2}
\end {equation}
so that the multiplier is
\begin {equation}
 {M_{12}} =  - {\displaystyle \frac {{u_1}}{(x {u_1}{u_2} +
x - {u_1}^{2})(x{u_2}
 -{u_1})}}. \label {2.3}
\end {equation}
If we integrate $M_{12}$ twice with respect to $u_2$, then from
formula (\ref{1.1}) we obtain the Lagrangian
\begin {eqnarray}
L_{12} & = & -\frac{u_1}{x^3}\left (xu_2 - u_1\right)\log ( x{u_2}
- u_1) +\frac{  xu_1{u_2} +x - u_1 ^{2 }}{x^3}\log ( xu_1{u_2} + x
- u_1 ^{2})
\nonumber\\
&&-\frac{1}{x^2}+{f_1}(x, u_1){u_2} + {f_2}(x, u_1), \label {2.4}
\end {eqnarray}
where $f_1 (x,u_1) $ and $f_2 (x,u_1) $ are arbitrary functions of
integration. If we substitute (\ref {2.4}) into the
Euler-Lagrangian equation, we obtain the constraint
\begin {equation}
{\frac {\partial {f_1} }{\partial x}}- {\frac {\partial {f_2}
}{\partial u_1}}
 = \frac{x-u_1^2}{x^3 u_1} \label {2.5}
\end {equation}
on the hitherto arbitrary functions $f_1 $ and $f_2 $. This
Lagrangian was not found by Ibragimov.

As  it was shown in \cite {Nucci 07 a}, \cite{Nucci 07 c}, $f_1,
f_2$ are related to the gauge function $g=g(x,u_1)$. In fact, we
may assume
\begin {equation}
f_1=  \frac{\partial g}{\partial u_1}, \quad\quad f_2=
\frac{\partial g}{\partial x} +\frac{2x\log(u_1)-
 u_1^2}{2x^3},
\label{gf1f2}
\end{equation}
namely the arbitrariness in the Lagrangian (\ref {2.4}) can be
expressed as a total time derivative. Such a Lagrangian has been
termed `gauge variant' \cite {Levy-Leblond 71 a} and is notable in
that the presence of the arbitrary function $g$ has no effect upon
the number of Noether point symmetries \cite {Nucci 07 a}.  In
this respect it could be regarded as part of the boundary term in
the way Noether put it in her formulation of her theorem \cite
{Noether 18 a}.  The class of Lagrangians described by (\ref
{2.4}) is an equivalence class.

  We
 observe that there are two singularities given by
\begin {equation}
x {u_1} {u_2} + x -  {u_1}^{2} = 0\quad\mbox {\rm and}\quad x
{u_2}
 -  {u_1} = 0 \label {2.6}
\end {equation}
When we solve these two equations, i.e.:
\begin{equation} y'=-\frac{1}{y}+\frac{y}{x}, \quad\quad {\rm and}
\quad y'=\frac{y}{x}, \label{sing1}
\end{equation} we recover the singular
solutions of (\ref {1.7}) associated with the singularities of the
Lagrangian (\ref{2.4}).

If we take the Lagrangian (\ref {2.4}) subject to (\ref {2.5}) and
calculate its Noether point symmetries, we find that there is a
single Noether point symmetry which is $\Gamma_2 $ in (\ref{1.9}).
The corresponding integral is
\begin {equation}
I = \frac {u_1} { xu_1u_2+x - u_1 ^2 }. \label {2.701}
\end {equation}
With an integral and a multiplier we can generate a second
multiplier by a reversal of the property that the quotient of two
multipliers is an integral.  The multiplier is just
\begin {equation}
M_{1} =\frac{M_{12}}{I}= - {\displaystyle \frac {1}{x {u_2} -
u_1}}. \label{2.8}
\end{equation}
Now we can calculate a second Lagrangian from the multiplier (\ref
{2.8}), and find
\begin{equation}
L_{1} = \frac{u_1-xu_2}{x^2}\log(u_1 -
xu_2)+\frac{u_2}{x}+f_1(x,u_1)u_2+f_2(x,u_1), \label{2.7}
\end{equation}
with the constraint:
\begin {equation}
{\frac {\partial {f_1} }{\partial x}}- {\frac {\partial {f_2}
}{\partial u_1}}
 = \frac{u_1^2+x}{x^2 u_1^2} \label {2.7b}
\end {equation}
or equally in terms of the gauge function $g=g(x,u_1)$
\begin {equation}
f_1=  \frac{\partial g}{\partial u_1}, \quad\quad f_2=
\frac{\partial g}{\partial x} +\frac{x-
 u_1^2}{u_1x^2}.\label {2.7c}
\end {equation}
 If we
take the Lagrangian $L_1$ in (\ref {2.7}) and calculate its
Noether point symmetries, we find that there is a single Noether
point symmetry which is $\Gamma_2 $ in (\ref{1.9}). The
corresponding integral is $I$ in (\ref{2.701}).

We can generate many (infinite) different Jacobi Last Multipliers
of equation (\ref{1.7}) and consequently many (infinite) different
Lagrangians. In fact we may take any function of the first
integral $I$ in (\ref{2.701}) and then its product with either
$M_{12}$ or $M_{1}$ will generate a new Jacobi Last Multiplier.
For example, we obtain the following multiplier:
\begin {equation}
M_{2} =M_{12}\frac{-2}{I^2}= 2 {\displaystyle \frac
{xu_1u_2+x-u_1^2}{u_1(x{u_2} - u_1)}}. \label{MI2}
\end{equation}
and consequently Lagrangian:
\begin {equation}
L_2=-2\frac{u_1-xu_2}{xu_1}\log(u_1-xu_2)+\frac{u_2(u_1u_2-2)}{u_1}+f_1(x,u_1)u_2+f_2(x,u_1),
\label{L2}
\end{equation}
with the constraint:
\begin {equation}
{\frac {\partial {f_1} }{\partial x}}- {\frac {\partial {f_2}
}{\partial u_1}}
 = -\frac{2}{ u_1^3}.\label{L2f}
\end {equation}
or equally in terms of the gauge function $g=g(x,u_1)$
\begin {equation}
f_1=  \frac{\partial g}{\partial u_1}, \quad\quad f_2=
\frac{\partial g}{\partial x} -\frac{1}{u_1^2}. \label{L2fb}
\end{equation}

 If we calculate the Noether point symmetries of the Lagrangian $L_2$ in (\ref {L2}), we find
that both $\Gamma_1$ and $\Gamma_2 $ in (\ref{1.9}) are Noether
point symmetries. The corresponding integrals are
\begin {equation}
I_1 = \log\left(\frac{u_1^2}{x}-u_1u_2\right) -
\frac{1}{u_1^2}\left(u_1^3u_2-xu_1^2u_2^2-2xu_1u_2-x\right),\label
{I1}
\end {equation}
 and
\begin {equation}
I_2 =\frac{1}{I^2}= \left(\frac { xu_1u_2+x - u_1 ^2
}{u_1}\right)^2 , \label {I2}
\end {equation}
respectively. It is  worth noting that both singular solutions
obtained in (\ref{sing1}) correspond to these integrals taking the
particular value of zero, namely, when each integral is a
configurational invariant \cite{Hall 83 a}, \cite{Sarlet 85 a}, we
obtain a singular solution.

One of the two Lagrangians derived by Ibragimov \cite{Ibragimov 99
a}[eq (42), p. 223] for equation (\ref{1.7}) is the following
\begin{equation}
L_{N1} =
\frac{1}{xu_1}+\left(\frac{u_1}{x^2}-\frac{u_2}{x}\right)\log\left(\frac{u_1^2}{x}-u_1u_2\right),
\label{N1L1}
\end{equation}
which is a particular case of the Lagrangian $L_1$ in (\ref{2.7})
 with
\begin {equation}
 f_1 = \frac{1}{x}(- \log(u_1) + \log(x) - 1), \quad f_2= \frac{u_1}{x^2}\left(\log(u_1)-
\log(x)\right) + \frac{1}{xu_1}.
\end{equation}

The other Lagrangian of Ibragimov \cite{Ibragimov 99 a}[eq (54),
p. 225] is the following
\begin {equation}
L_{N2} =
-\frac{1}{u_1^2}+\frac{u_1^2}{x^2}-2\frac{u_1u_2}{x}+u_2^2
-2\left(\frac{1}{x}-\frac{u_2}{u_1}\right)\log\left(\frac{u_1^2}{x}-u_1u_2\right),
\label{N1L2}
\end{equation} which is a particular case of the  Lagrangian $L_2$ in (\ref{L2})
with
\begin {eqnarray}
 f_1 &=& -\frac{2}{xu_1}\left(u_1^2- x\log(u_1) +x \log(x) - x\right),
 \nonumber\\ f_2&=& -\frac{1}{x^2u_1^2}\left(x^2-u_1^4+2xu_1^2\log(u_1)-
2xu_1^2\log(x)\right).
\end{eqnarray}

\section {Jacobi Last Multipliers and Lagrangians for (\ref {1.8})}

The system of first-order differential equations corresponding to (\ref {1.8}) is
\begin {eqnarray}
& & {u_1'} =  {u_2} \nonumber \\
& & {u_2'} =  - {\displaystyle \frac { {u_2}}{x}}  + \e^{ {u_1}}.
\label {3.1}
\end {eqnarray}
In this case the application of formula (\ref {12.5}) or
equivalently (\ref{Meq})
 does produce a multiplier.  It is
\begin {equation}
M_0 = x \label {3.2}
\end {equation}
from which we obtain the Lagrangian
\begin {equation}
{L_0} = {\displaystyle \frac {x{u_2}^{2}}{2}}  + {f_1}(x,
u_1){u_2} + {f_2}(x, u_1) \label {3.3}
\end {equation}
with the constraint on the two functions of integration being
\begin {equation}
\frac {\partial f_1 }{\partial x} - \frac {\partial f_2 }{\partial
u_1} =-x\e^{u_1},
 \label {3.301}
\end {equation}
or equally in terms of the gauge function $g=g(x,u_1)$
\begin {equation}
f_1=  \frac{\partial g}{\partial u_1}, \quad\quad f_2=
\frac{\partial g}{\partial x}+x\e^{u_1}. \label{3.3b}
\end{equation}
This Lagrangian admits one Noether's symmetry namely $\Sigma_1$ in
(\ref{1.10}) and yields the following first integral:
\begin {equation}
 {I_0} =   4x {u_2} +  {u_2}^{2}x^{2 } - 2\e^{ {u_1}}x^{2}.  \label {3.8}
\end {equation}

We use the two symmetries $\Sigma_1, \Sigma_2$ in (\ref{1.10}) and
the vector field of the system (\ref {3.1})
 to obtain a second multiplier.  The matrix is
\begin {equation}
 {{\rm Mat}_{12}} =  \left[ {\begin{array}{ccc} 1 &  {u_2} &
-
{\displaystyle \frac { {u_2}}{x}}  + \e^{ {u_1}} \\
[2ex] x {\log}(x) & \quad - 2\left(1+  {\log}(x)\right) &  \quad -
{\displaystyle\frac {2}{x}}  -  {u_2}\left(1 +  {\log}(x)\right)
\\ [2ex] x & -2 &  -  {u_2}
\end{array}}
 \right] \label {3.4}
\end {equation}
and the corresponding multiplier is
\begin {equation}
 {M_{12}} =  - {\displaystyle \frac {x}{4 + 4x {u_2} +
 {u_2}^{2}x^{2} - 2\e^{ {u_1}}x^{2}}}. \label {3.5}
\end {equation}
 Thus formula (\ref{1.1}) yields the following  Lagrangian
\begin {eqnarray}
L_{12} &=& \frac{1}{x}\log\left(-xu_2-2-\sqrt{2}x\e^{u_1/2}\right)
-\frac{1}{2x}\log\left(\frac{xu_2+2+\sqrt{2}x\e^{u_1/2}}{xu_2+2-\sqrt{2}x\e^{u_1/2}}\right)
\nonumber\\&&+\frac{\sqrt{2}(xu_2+2)}{4x^2\e^{u_1/2}}
\log\left(\frac{xu_2+2+\sqrt{2}x\e^{u_1/2}}{xu_2+2-\sqrt{2}x\e^{u_1/2}}\right)
+
 {f_1} {u_2} +  {f_2} \label {3.6}
\end {eqnarray}
with the constraint
\begin {equation}
\frac {\partial f_1}{\partial x} - \frac {\partial f_2}{\partial
u}  = 0. \label {3.7}
\end {equation}
or equally  in terms of the gauge function $g=g(x,u_1)$
\begin {equation}
f_1=  \frac{\partial g}{\partial u_1}, \quad \quad f_2=
\frac{\partial g}{\partial x}. \label{3.7b}
\end{equation}
In a curious repetition of the situation with (\ref {1.7}) we find
that the two Lagrangians $L_0$ in (\ref{3.3}) and $L_{12}$ in
(\ref{3.6}) have the same Noether point symmetry $\Sigma_1 $ in
(\ref{1.10}) and lead to what is functionally the same integral
$I_0$ in (\ref{3.8}). Both Lagrangians were not found by
Ibragimov. Indeed Ibragimov did not look for Lagrangians of
equation (\ref {1.8}) admitting one Noether point symmetry.

Since we have two multipliers, we can obtained a first integral
given by their ratio, namely
\begin {equation}
 \frac{M_0}{M_{12}} =  - (xu_2+2)^2 + 2\e^{ {u_1}}x^{2}=-I_0-4.  \label {3.8b}
\end {equation}

We can generate many (infinite) different Jacobi Last Multipliers
of equation (\ref{1.8}) and consequently many (infinite) different
Lagrangians. In fact we may take any function of the first
integral $I_0$ in (\ref{3.8}) and then its product with either
$M_{12}$ or $M_{0}$ will generate a new Jacobi Last Multiplier.
For example, we obtain the following multiplier:
\begin {equation}
M_{2} =-\frac{M_{0}}{\sqrt{I_0+4}}= - {\displaystyle \frac
{x}{\sqrt{(xu_2+2)^2 - 2\e^{u_1}x^{2}}}}. \label{M2_2}
\end{equation}
and consequently the following Lagrangian:
\begin {eqnarray}
L_2&=&
-\left(u_2+\frac{2}{x}\right)\log\left(\frac{\sqrt{(xu_2+2)^2 -
2\e^{
{u_1}}x^{2}}+xu_2+2}{\sqrt{2}x\e^{u_1/2}}\right)\nonumber\\
&&+\sqrt{(xu_2+2)^2 - 2\e^{u_1}x^{2}}+f_1(x,u_1)u_2+f_2(x,u_1),
\label{L2_2}
\end{eqnarray}
with  either the constraint (\ref{3.7}) or (\ref{3.7b}).
 This Lagrangian admits
 both $\Sigma_1$ and $\Sigma_2 $ in (\ref{1.10}) as Noether point
symmetries, and the corresponding integrals are
\begin {equation}
I_1 =\sqrt{I_0+4}=\sqrt{(xu_2+2)^2 - 2\e^{u_1}x^{2}} ,\label
{I1_2}
\end {equation}
 and
\begin {equation}
I_2 =\sqrt{(xu_2+2)^2 - 2\e^{
{u_1}}x^{2}}\log(x)+2\log\left(\frac{\sqrt{(xu_2+2)^2 - 2\e^{
{u_1}}x^{2}}+xu_2+2}{\sqrt{2}x\e^{u_1/2}}\right), \label {I2_2}
\end {equation}
respectively.

The Lagrangian of Ibragimov \cite{Ibragimov 99 a}[eq (90), p. 234]
is  a particular case of the  Lagrangian $L_2$ in (\ref{L2_2})
with the
 gauge function equal to zero\footnote{In \cite{Ibragimov 99 a} there are some
 missprints.}.

\strut\hfill

 The last multiplier $M_{12} $ in (\ref{3.5}) becomes
singular if
\begin {equation}
y' = -\frac {2} {x} \pm\sqrt {2}\e ^ {y/2}. \label {3.9}
\end {equation}
Equation (\ref {1.8}) is satisfied by each of the first-order equations
 in (\ref {3.9}) and so we obtain the two singular solutions
\begin {equation}
y = x\left (C \mp\sqrt {2}x\right) \label {3.10}
\end {equation}
thereby supplementing the results given in \cite {Ibragimov 05 a}.

\section {Final Remarks}
When one seeks a Lagrangian of an elementary equation, it is
usually possible to guess the form of at least one Lagrangian.  In
the case of not so elementary equations an approach using
guesswork is likely to lead to frustration.  Consequently any
development which can replace guesswork or intuition by a
well-defined procedure is to be welcomed.  Usually there is a
price to pay for the elimination of guesswork\footnote {Naturally
it can be argued that guesswork is a process of looking for some
type of symmetry in an intuitive matter.}. In this paper we have
considered two test equations proposed by Ibragimov to illustrate
his concept of the use of invariant Lagrangians to provide a new
method for the integration of nonlinear equations.  We have
demonstrated that some quite old knowledge is available for a
successful resolution of the same problems.  The combination of
the concept introduced by Jacobi in his last multiplier, the
application by Lie of his ideas of invariance under the
transformations generated by continuous groups and the
specialisation to the Action Integral by Noether provides us with
a very powerful and simple tool for the resolution of ordinary
differential equations which possess a reasonable amount of
symmetry.  We have seen in the two examples considered here that
they provide richer results when considered from a more classical
viewpoint. The Jacobi last multiplier yields more general
Lagrangians than those find by Ibragimov, and many more can be
generated. One could consider that the combination of Jacobi and
Lie gives sufficient material to deal with these equations. In
that sense it could be argued that the theorem of Noether is
already implicit in the work of Jacobi and Lie. However, we did
see that further results were available to us by an application of
Noether's Theorem to the information already obtained.

\section*{Acknowledgements}

This work was undertaken while PGLL was enjoying the hospitality
of Professor MC Nucci and the facilities of the Dipartimento di
Matematica e Informatica, Universit\`a di Perugia.  The continued
support of the University of KwaZulu-Natal is gratefully
acknowledged.

\end{document}